\documentclass[12pt]{article}
\usepackage{amssymb}
\usepackage[dvips]{graphicx}
\usepackage{epsfig}
\renewcommand{\arraystretch}{2}
\begin{document}
\setlength{\baselineskip}{0.75cm}
\setlength{\parskip}{0.45cm}
\begin{titlepage}
\begin{flushright}
DO-TH 97/26 \linebreak
November 1997
\end{flushright}
\vskip 1.2in
\begin{center}
{\Large\bf Leading log radiative corrections to polarized heavy flavour 
leptoproduction}
\vskip 0.8in
{\large I.\ Schienbein}
\end{center}
\vskip .3in
\begin{center}
{\large Institut f\"{u}r Physik, Universit\"{a}t Dortmund \\
D-44221 Dortmund, Germany}
\end{center}
\vskip 1in
{\large{\underline{Abstract}}}

\noindent
${\cal O}(\alpha)$ QED radiative corrections to polarized neutral current  
leptoproduction of heavy quark flavours are calculated in the leading log 
approximation for HERMES, COMPASS, and HERA kinematics.
Besides the inclusive case, we derive general $z_h$-differential formulae for
the polarized boson gluon fusion process and use them to calculate radiative 
corrections to semi-inclusive polarized heavy flavour production 
in the case of HERA kinematics.
\end{titlepage}
\section{Introduction}
So far, the polarized gluon density $\Delta g(x,Q^2)$ is nearly unknown, see
e.g.~\cite{gehrmann97}.
In order to improve this status several experiments are designed to measure
polarized heavy flavour production which is directly linked to $\Delta g(x,Q^2)$
already in leading order (LO) QCD via the boson (photon) gluon fusion mechanism.
Such measurements will be performed by, e.g., the HERMES Collaboration with
its charm detector upgrade \cite{hermesc} and the COMPASS Collaboration at CERN
\cite{compass}.
Using fixed targets these efforts are restricted to rather low energies.
A polarized HERA collider mode \cite{polhera} would allow for measurements
in a kinematical region not accessible to
any other polarized experiment \cite{ball}.

Radiative corrections to polarized deep inelastic scattering using
HERA and HERMES kinematics have been recently calculated in complete
$\cal{O}(\alpha)$ and in $\cal{O}(\alpha)$ leading log approximation (LLA) \cite{bb96}
disregarding heavy flavour production.
However, at small $x\lesssim 10^{-3}$ the heavy (charm) quark contribution
$g_1^c(x,Q^2)$ to the spin structure function $g_1^P(x,Q^2)$ becomes relevant.
Keeping in mind that heavy flavour production processes are semi-inclusive it
is recommendable to perform the radiative corrections on a more differential
level. As in the unpolarized case \cite{ijs97} we study the momentum 
($z_D=p \cdot p_D / p \cdot q$)
distributions of the $D$-mesons produced in such events.
For this purpose we derive the $z_h$-differential expressions
($z_h=p \cdot p_h / p \cdot q$) with general couplings, where
$h$ denotes a heavy (charm) quark, 
for the polarized boson gluon fusion process. This allows us to study electroweak
(ew) radiative effects in heavy flavour production. Such effects have been recently 
investigated in the
case of the light flavours $u$, $d$, $s$ \cite{bb96,ais97} and there turned out 
to be sizeable at small Bjorken-$x$ ($x=10^{-3}$, $10^{-2}$) \cite{bb96}.

Thus, in this paper we calculate radiative corrections to 
inclusive and semi-inclusive polarized heavy flavour 
production in LLA for HERMES, COMPASS, and HERA
kinematics. 
Since in leading order QED the polarized and unpolarized splitting functions
are the same, i.e. $\Delta P_{ee}(x)=P_{ee}(x)$, 
the whole formalism can be adopted from the unpolarized case \cite{ijs97} merely
replacing the unpolarized Born cross sections by the corresponding 
polarized expressions.

The paper is organized as follows: In Sect.~2 the Born-level expressions for 
polarized deep inelastic scattering are given and the general $z_h$-differential
expressions for polarized heavy flavour production via the booson gluon fusion
mechanism are derived. In Sect.~3 we provide the theoretical framework 
to calculate inclusive and semi-inclusive  $\cal{O}(\alpha)$-QED corrections 
in LLA, followed by numerical results for the fully massive 
radiative corrections to heavy flavour production in Sect.~4.
Finally, we summarize our results in Sect.~5.
\section{Born cross section}
In our calculations we will restrict on the longitudinal spin asymmetry
$d\Delta \sigma=(d\sigma(\lambda,S_L)-d\sigma(\lambda,-S_L))/2$, where
$d\sigma(\lambda,S_L)$ denotes the differential Born cross section for deep
inelastic scattering of longitudinally polarized leptons 
off longitudinally polarized nucleons.
For definiteness, we consider the process
$l(k_1,\lambda)P(p,S_L)\rightarrow l^\prime(k_2)X(p_X)$ where 
$\lambda$ and $k_1$
are the helicity and the 4-momentum of the incoming lepton, 
$S_L$ and $p$ the spin and the 4-momentum of the nucleon and
$k_2$ the $4$-momentum  of the final lepton.
The Born cross section can be written as a contraction of a leptonic tensor
$L_{\mu\nu}$ with a hadronic tensor $W^{\mu\nu}$: 
\begin{eqnarray}
d\sigma^{NC}&=&d\sigma^{\gamma \gamma}+2 Re (d\sigma^{\gamma Z})+d\sigma^{ZZ}
\nonumber \\
d\sigma^{CC}&=&d\sigma^{WW}
\nonumber \\
d\sigma^{BB^{\prime}}&=&\frac{2\pi\alpha^{2}}{Q^4}y\ 
\chi_{B}\chi_{B^{\prime}}\left(L_{\mu\nu}^{B B^\prime}
W_{B B^\prime}^{\mu\nu}\right) dx dy,
\label{dsb1b2}
\end{eqnarray}
with $B,B^\prime =\gamma,Z$ or $W$, $\alpha$ the QED coupling constant,
and $x$, $y$, $Q^2$ the well-known variables in deep inelastic scattering (DIS).
$\chi_{B}$ collects some constants and the ratio of $B$-propagator 
and photon-propagator:
\begin{eqnarray}
\chi_{\gamma}(Q^2)&=&1
\nonumber\\
\chi_{Z}(Q^2)&=&\frac{1}{(2 \sin \theta_W \cos \theta_W)^{2}}
\frac{Q^{2}}{Q^{2}+M_{Z}^{2}}
\\
\chi_{W}(Q^2)&=&\frac{1}{8 (\sin {\theta_W})^2} \frac{Q^2}{Q^2+M_{W}^2}\ ,
\nonumber
\label{chi}
\end{eqnarray}
where $\theta_W$ is the  Weinberg angle and $M_Z$, $M_W$ 
are the masses of the $Z^0$ and $W$ boson, respectively.

Using a general $\gamma_{\mu}(V-A\gamma_5)$-structure
for the leptonic current, 
$L_{\mu\nu}^{B B^\prime}$ reads:
\begin{eqnarray}
L_{\mu\nu}^{B B^\prime}(\lambda)&=&\sum_{\lambda '}
\bar{u}(\lambda,k_1)\Gamma^{B^\prime} _{\nu}u(\lambda ',k_2)\bar{u}(\lambda ',k_2)
\Gamma^{B}_{\mu}u(\lambda,k_1)
\nonumber \\
&=&
2C_l^{BB^\prime}
(k_{1\mu}k_{2\nu}+k_{1\nu}k_{2\mu}-(k_1\cdot k_2)g_{\mu\nu}
+2 i \lambda\epsilon_{\mu \nu \rho \sigma} k_1^{\rho}{k_2} ^{\sigma})\ .
\label{lb1b2}
\end{eqnarray}
$\Gamma_{\mu}^{B}=\gamma_{\mu}(V_{l}^{B}-A_{l}^{B}\gamma_5)$ and
$C_l^{BB^\prime}=(V_l^B-\lambda p_l A_l^B)(V_l^{B^\prime}-\lambda p_l A_l^{B^\prime})$
are built from the vector (V) and axialvector (A) couplings of the fermion-gauge
boson vertex $\gamma_{\mu}(V_f^B-A_f^B\gamma_5)$. They are given by
\begin{center}
\renewcommand{\arraystretch}{1.2}
\begin{tabular}{l|ll}
$B$&$V_{f}^B$&$A_{f}^B$\\ \hline
$\gamma$&$Q_f$&$0$\\
$Z$&$I_{f}^{3}-2 Q_f \sin^2\theta_{w}$&$I_{f}^{3}$\\
$W$&$1$&$1$
\end{tabular}
\end{center}
where $Q_f$ ($Q_{e^-}=-1$,$Q_{e^+}=+1$,$Q_{u,c,t}=+2/3$,$Q_{d,s,b}=-1/3$)  
and $I_{f}^{3}$ ($I_{e^+,u,c,t}^{3}=1/2$, $I_{e^-,d,s,b}^{3}=-1/2$)
are the charge and the isospin of the fermion.
$p_l$ accounts for the fact that the result for $e^+p$-scattering can be obtained
from the $e^-p$-results by the replacement $A_l^B \to -A_l^B$, i.e. $p_{e^+}=-1$ and
$p_{e^-}=1$.
Furthermore, it is convenient to define two further combinations of the 
couplings at this place:
\begin{eqnarray}
S_{f,\pm}^{BB^\prime}&=&{V_f}^B {V_f}^{B^\prime} \pm {A_f}^B {A_f}^{B^\prime},
\nonumber\\
R_{f,\pm}^{BB^\prime}&=&({V_f}^B {A_f}^{B^\prime}\pm {V_f}^{B^\prime} {A_f}^B)/2\,
\end{eqnarray}
such that $C_l=S_{l,+}-p_l\lambda 2R_{l,+}$.

Assuming current conservation, the polarized part $W_{\mu\nu}(p,q,S)$ 
of the hadronic tensor depends in general on five structure 
functions $g_i(x,Q^2)$.
In the literature different definitions for the structure functions are used.
For a comparison of several conventions see \cite{bk}.
We use the following form:
\begin{eqnarray}
W_{\mu\nu}^{BB^\prime}(p,q,S)&=&i\epsilon_{\mu\nu\lambda\sigma}
\frac{q^\lambda S^\sigma}{p\cdot q}g_1^{BB^\prime}
+i\epsilon_{\mu\nu\lambda\sigma}\frac{q^\lambda(p\cdot q\ S^\sigma
-S\cdot q\ p^\sigma)}
{(p\cdot q)^2}g_2^{BB^\prime}
\\
&&+\left(g_{\mu\nu}-\frac{q_{\mu}q_{\nu}}{q^2}\right)
\frac{S\cdot q}{p\cdot q}g_3^{BB^\prime}
-\left(\frac{S\cdot q}{2(p\cdot q)^2}p_{\mu}^\perp p_{\nu}^\perp+
\frac{p_{\mu}^\perp S_{\nu}^\perp+
p_{\nu}^\perp S_{\mu}^\perp}{4p\cdot q}\right)g_4^{BB^\prime}
\nonumber\\
&&+\left(\frac{p_{\mu}^\perp S_{\nu}^\perp+
p_{\nu}^\perp S_{\mu}^\perp}{4p\cdot q}
-\frac{S\cdot q}{2(p\cdot q)^2}p_{\mu}^\perp p_{\nu}^\perp\right)g_5^{BB^\prime}
\nonumber
\end{eqnarray}
with
\begin{displaymath}
p_{\mu}^\perp=p_{\mu}-\frac{p\cdot q}{q^2}q_{\mu},\qquad
S_{\mu}^\perp=S_{\mu}-\frac{S\cdot q}{q^2}q_{\mu}\ .
\end{displaymath}
For longitudinally polarized nucleons
the nucleon spin vector can be chosen in the nucleon rest frame 
to be $S_L=(0,0,0,M)$.
Neglecting terms of the order ${\cal O}(M^2/S)$ (S=CMS-energy) one can replace 
$S_L$ by $-\lambda_N p$ where $\lambda_N$ is the degree of
the nucleon polarization \cite{manohar} and discard $g_2$ and $g_5$.
This yields
\begin{eqnarray}
W_{\mu\nu}&=&\lambda_N\left(-i\epsilon_{\mu\nu\lambda\sigma}
\frac{q^\lambda p^\sigma}{p\cdot q}g_1-
\left(g_{\mu\nu}-\frac{q_{\mu}q_{\nu}}{q^2}\right)g_3
+\frac{p_{\mu}^\perp p_{\nu}^\perp}{p\cdot q}g_4\right)
\label{Wb1b2}
\end{eqnarray}
in agreement with the hadronic Tensor given in \cite{vw91}.
Using eq.~(\ref{dsb1b2}), (\ref{lb1b2}) and (\ref{Wb1b2}) one finds the 
following result for the
longitudinal spin asymmetry (polarized part of the Born cross section):
\begin{eqnarray}
\frac{d^2\Delta \sigma}{dx dy}(l(\lambda)P(\lambda_N)\rightarrow l'X)&=&
\frac{4\pi \alpha^2}{Q^2xy}\lambda_N \Big[y(2-y)xG_1(x,Q^2)+y^2 x G_3(x,Q^2)
\nonumber\\
&&\qquad \qquad +(1-y)G_4(x,Q^2) \Big]
\label{polborn}
\end{eqnarray}
with the structure functions
\begin{eqnarray}
G_1(x,Q^2)&=&\lambda\sum_{B,B^\prime} C_l^{BB^\prime}\chi_B(Q^2)\chi_{B^\prime}(Q^2)
g_1^{BB^\prime}(x,Q^2)
\nonumber\\
G_{3,4}(x,Q^2)&=&\sum_{B,B^\prime} C_l^{BB^\prime}\chi_B(Q^2)\chi_{B^\prime}(Q^2)
g_{3,4}^{BB^\prime}(x,Q^2)\ .
\label{struct}
\end{eqnarray}
%
%

The structure functions $g_i^{BB^\prime}$ can be calculated within
the quark parton model.
The two relevant partonic processes are in leading order (LO) the polarized 
heavy flavour production via the boson-gluon fusion subprocess
$B^*(q)g(k)\rightarrow \bar{Q}_1(p_1)Q_2(p_2)$ and the electron-quark scattering
$B^*q\rightarrow q$.
\subsection{Polarized heavy flavour production}
The structure functions $g_{i=1,3,4}^h$ for a pair of heavy quarks $Q_{1,2}$ with 
masses $m_{1,2}$ can be obtained by a convolution
of the polarized gluon density $\Delta g$ with the polarized structure 
functions $\hat{g}^h_{i=1,3,4}$ of the partonic subprocess:
\begin{eqnarray}
g_{1,3}^{h,BB^\prime}(x,Q^2,\mu^2)=
\int_{ax}^1\ \frac{dz_2}{z_2}\ \Delta g(z_2,\mu^2)
\ \hat{g}_{1,3}^{h,BB^\prime}(x/z_2,Q^2,\mu^2),
\nonumber\\
g_{4}^{h,BB^\prime}(x,Q^2,\mu^2)=
\int_{ax}^1\ \frac{dz_2}{z_2}\ z_2 \Delta g(z_2,\mu^2)
\ \hat{g}_{4}^{h,BB^\prime}(x/z_2,Q^2,\mu^2),
\label{hstruct}
\end{eqnarray} 
with $a=1+(m_1+m_2)^2/Q^2$.
As usual we identify the scales $\mu_R=\mu_F=:\mu$ in $\Delta g(z_2,\mu_F)$
and $\alpha_s(\mu_R)$ (see below).

Following the general framework given in ref.~\cite{vw91},
we calculate the polarized partonic structure functions 
$\hat{g}^{h,BB^\prime}_{i=1,3,4}$ in $z_{h}$-differential
form ($h=c$, $b$, $t$) by writing the two particle phase space as 
$dR_2=dz_{h}/(8\pi)$ with
$z_{h}=k\cdot p_2/k\cdot q$.
The general $z_{h}$-dependence of the polarized structure functions turns out
to be the same as in the unpolarized case \cite{ks97}:
\begin{equation}
\hat{g}_{i=1,3,4}^{h,BB^\prime}(x,Q^2,\mu^2;z_{h})
=\frac{\alpha_s(\mu^2)}{4\pi}\left[
\frac{{A_i}^{BB^\prime}}{(1-z_h)^2}
+\frac{{B_i}^{BB^\prime}}{z_h^2}
+\frac{{C_i}^{BB^\prime}}{(1-z_h)}
+\frac{{D_i}^{BB^\prime}}{z_h}
+{E_i}^{BB^\prime}
\right]
\label{generalg}
\end{equation}
with (suppressing the indices $BB^{\prime}$)
\begin{eqnarray*}
A_1(x,Q^2)&=&S_{q,+}\frac{m_1^2}{Q^2}x 
\\
C_1(x,Q^2)&=&S_{q,+}\left[\Delta P_{qg}(x)+\frac{\Delta m^2}{Q^2}x\right]
\\
E_1(x,Q^2)&=&S_{q,+}(-2\Delta P_{qg}(x))
\\
A_3(x,Q^2)&=&-2\frac{R_{q,+}}{S_{q,+}}A_1(x,Q^2)
\\
C_3(x,Q^2)&=&-2\frac{R_{q,+}}{S_{q,+}}C_1(x,Q^2)
\\
E_3(x,Q^2)&=&0
\\
A_4(x,Q^2)&=&R_{q,+}\left[-4x^2\frac{m_1^2}{Q^2}\left
(1+\frac{\Delta m^2}{Q^2}\right)\right]
\\
C_4(x,Q^2)&=&R_{q,+}2x\left[-2\Delta P_{qg}(x)+
\frac{\Delta m^2}{Q^2}(1-4x)-2x\frac{\Delta m^4}{Q^4}\right]
\\
E_4(x,Q^2)&=&0
\\
B_{i={1 \atop 3,4}}\left(x,Q^2\right)&=&\pm A_i\left(x,Q^2\right)
[m_1\leftrightarrow m_2]
\\
D_{i={1 \atop 3,4}}\left(x,Q^2\right)&=&\pm C_i\left(x,Q^2\right)
[m_1\leftrightarrow m_2]
\end{eqnarray*}
where $\Delta m^n \equiv m_2^n-m_1^n\ ,\ m_{1,2}$ being the mass of
the heavy quark $Q_{1,2}$ and $\Delta P_{qg}(x)=(2x-1)/2$ the polarized
$g\to q$ splitting function.

The $z_{h}$-integrated results in \cite{vw91} are easily recovered if one 
notices that $1/(1-z_{h})\rightarrow L=\ln (1-z_{h,min})/(1-z_{h,max})$,
$1/z_{h}\rightarrow \tilde{L}=\ln z_{h,max}/z_{h,min}
=L[m_1\leftrightarrow m_2]$, 
$1/(1-z_{h})^2\rightarrow v \bar{v}\ W^2/m_1^2$ and
$1/z_{h}^2\rightarrow v \bar{v}\ W^2/m_2^2$ with the $z_{h}$-integration 
boundaries
\cite{schuler88}
\begin{equation}
z_{h_{min,max}}(x,Q^2)=\frac{1}{2}\left[1+\Delta m^2/W^2\pm v\bar{v}\right]
\label{zhbound}
\end{equation} 
where 
$v^2(x,Q^2)=1-(m_1+m_2)^2/W^2$,
${\bar{v}}^2(x,Q^2)=1-(m_1-m_2)^2/W^2$ and
$\displaystyle W^2(x,Q^2)=\frac{Q^2}{x}(1-x)$.
In the following we are only interested in neutral current charm production, i.e.,
$m_1=m_2=m_c$, $\bar{v}=1$ and $\tilde{L}=L=\ln (1+v)/(1-v)$. 
In this case the $z_{h}$-integrated partonic structure functions read:
\begin{eqnarray}
\hat{g}_1^{c,BB^\prime}(x,Q^2,\mu^2)&=&\frac{\alpha_s(\mu^2)}{4\pi}
S_{q,+}^{BB^\prime}\Big
[2\Delta P_{qg}(x)L+v(3-4x)\Big]
\nonumber\\
\hat{g}_3^{c,BB^\prime}(x,Q^2,\mu^2)&=&\hat{g}_4^{c,BB^\prime}(x,Q^2,\mu^2)=0\ .
\label{g3sub}
\end{eqnarray}
\subsection{Electron-quark scattering}
For completeness we list the LO neutral current results for 
electron quark scattering \cite{vw91}\footnote{Note that 
$R_{q,+}=-R_q/2$ with $R_q$ defined in \protect\cite{vw91}.}: 
\begin{eqnarray}
\label{udsg}
g_1^{BB^\prime}(x,Q^2)&=&\frac{1}{2}\sum_{q=u,d,s} S_{q,+}^{BB^\prime}[\Delta q(x,Q^2)
+\Delta \bar{q}(x,Q^2)]
\nonumber\\
g_3^{BB^\prime}(x,Q^2)&=&-\sum_{q=u,d,s} R_{q,+}^{BB^\prime}[\Delta q(x,Q^2)
-\Delta \bar{q}(x,Q^2)]
\\
g_4^{BB^\prime}(x,Q^2)&=&2xg_3^{BB^\prime}(x,Q^2)\ .
\nonumber
\end{eqnarray}
\section{Radiative corrections in LLA}
As already mentioned, in ${\cal O}(\alpha)$-LLA the structure of the QED-corrections 
to polarized DIS
is the same as in the unpolarized case, because
$\displaystyle \Delta P_{ee}(x)=P_{ee}(x)=\left[\frac{1+x^2}{1-x}\right]_{+}$.
We therefore restrict ourselves to give the formulae that we have used in 
this work.
For details we refer the reader to the literature \cite{ijs97,lla}.

We are using leptonic variables
\begin{eqnarray}
\label{outerkin}
q_{l}\equiv p_e-{p^{\prime}_e},\qquad Q_{l}^{2}\equiv -q_{l}^{2},
\qquad y_{l}
=\frac{p\cdot q_{l}}{p\cdot p_e},
\qquad x_{l}\equiv \frac{Q_{l}^{2}}{2p\cdot q_{l}}
=\frac{Q_l^2}{S y_l}\ ,
\end{eqnarray}
which are reconstructed by the $4$-momentum $p_e^{\prime}$ of the observed
electron and the $4$-momenta $p_e$, $p$ of the initial state electron and proton.
As usual $S=(p+p_e)^2$ is the center-of-mass (CMS) energy.

To describe semi-inclusive processes we further define the Lorentz-invariant
variable
\begin{equation}
z_D\equiv z_{D,l}=\frac{p\cdot p_D}{p\cdot q_l}\ 
\end{equation}
which is built from the $4$-momentum $p_D$ of the produced $D$-meson.
In the rest frame of the proton $z_D$ is the energy fraction carried by the 
$D$-meson with respect to the energy of the boson. 

The rescaled variables for the partonic subprocess like ,e.g.,
\newline $e(\hat{p}_e=z_1 p_e)+g(\hat{p}=z_2 p)\longrightarrow e(\hat{p}^\prime_e
=p_{e}^{\prime}/z_3)+c(p_c=P_D/z_4)+X$
read:
\begin{eqnarray}
\label{subkin}
\hat{q}&=&{\hat{p}_e}-{{\hat{p}^\prime_e}},\qquad 
\hat{Q}^{2}=-\hat{q}^{2}=\frac{z_1}{z_3}Q_l^2,\qquad
\hat{y}=\frac{\hat{p}\hat{q}}{\hat{p}{\hat{p}_e}}
=\frac{z_1z_3-1+y_l}{z_1z_3},
\nonumber\\
\hat{x}&=&\frac{\hat{Q}^{2}}{2\hat{p}\cdot \hat{q}}
=\frac{\hat{Q}^2}{\hat{S} \hat{y}}
=\frac{z_1x_ly_l}{z_2(z_1z_3-1+y_l)},\qquad
\hat{S}=(\hat{p}_e +\hat{p})^2=z_1 z_2 S,
\\
\hat{z}_c&=&\hat{p}\cdot p_c/\hat{p}\cdot \hat{q}=\frac{z_D y_l z_3}
{z_4(z_1z_3+y_l-1)}
\equiv \frac{z_D}{z_4}r
\nonumber
\end{eqnarray}
with $r$ defined by 
\begin{equation}
r\equiv \frac{y_l\ z_3}{z_1z_3+y_l-1}.
\label{r}
\end{equation}
The index $l$ to denote leptonic variables will be suppressed from now on.

\subsection{Inclusive case}
The corrections due to initial state radiation (ISR)
and final state radiation (FSR) can be written as \cite{ijs97}
\begin{eqnarray}
\frac{d^2\Delta\sigma^i}{dxdy}& = & \frac{\alpha L_{e}}{2\pi}
\int^1_{z_{1}^{min}}dz_1\left[\frac{1+z_1^2}{1-z_1}
(\Delta\sigma_0(z_1,1)-\Delta\sigma_0(1,1))\right] \nonumber
\\
&& +\ \frac{\alpha L_{e}}{2\pi} H(z_{1}^{min})\Delta\sigma_0(1,1),
\label{initial}
\end{eqnarray}
\begin{eqnarray}
\frac{d^2\Delta\sigma^f}{dxdy} & = & \frac{\alpha L_{e}}{2\pi}
\int^1_{z_{3}^{min}}dz_3\left[\frac{1+z_3^2}{1-z_3}
(\Delta\sigma_0(1,z_3)-\Delta\sigma_0(1,1))\right] \nonumber
\\
&& +\ \frac{\alpha L_{e}}{2\pi} H(z_{3}^{min})\Delta\sigma_0(1,1),
\label{final}
\end{eqnarray}
with $\alpha$ the electromagnetic coupling, $\displaystyle L_e=\ln Q^2/m_e^2$
($m_e=$ electron mass) and 
\begin{eqnarray}
\nonumber
H(z) & = & -\int_0^z dx \frac{1+x^2}{1-x}=2\ln(1-z)+z+\frac{1}{2}z^2\ .
\end{eqnarray}
$\Delta \sigma_0$ depends on the hard scattering process.

\noindent 1. Boson gluon fusion (heavy flavour production):
\begin{eqnarray}
\label{pgfs0}
\Delta\sigma_0(z_1,z_3) & = &\int^1_{z_{2}^{min}}dz_2\  
\Delta g\left(z_2,\mu ^2\right)
\ \frac{\partial{(\hat{x},\hat{y})}}{\partial{(x,y)}}
\ \frac{d^2\Delta\hat{\sigma}}{d\hat{x}d\hat{y}}, 
\end{eqnarray}
with  the partonic cross section
\begin{eqnarray}
\frac{d^2\Delta \hat{\sigma}}{d\hat{x} d\hat{y}}&=&
\frac{4\pi \alpha^2}{\hat{S}\hat{x}^2\hat{y}^2}\lambda_N 
\Big[\hat{y}(2-\hat{y})\hat{x}
\hat{G}_1(\hat{x},\hat{Q}^2)
\Big]
\label{partonic}
\end{eqnarray}
where
\begin{eqnarray}
\hat{G}_1(\hat{x},\hat{Q}^2)&=&\lambda\sum_{B,B^\prime} C_l^{BB^\prime}
\chi_B(\hat{Q}^2)\chi_{B^\prime}(\hat{Q}^2)
\hat{g}_1^{BB^\prime}(\hat{x},{Q}^2)
\label{substruct}
\end{eqnarray}
(($\hat{g}_1(\hat{x},\hat{Q}^2)$ from eq.~(\ref{g3sub}))
and the Jacobian
\begin{equation}
 \frac{\partial{(\hat{x},\hat{y})}}{\partial{(x,y)}}
=\frac{y}{z_2z_3(z_1z_3-1+y)}.
\label{jacobi1}
\end{equation}
The integration bounds are given by:
\begin{eqnarray}
\label{bounds}
z_{1}^{min} & = & \frac{1-y}{1-xy}+\frac{4m_c^2}{S(1-xy)},
\nonumber\\
z_{3}^{min} & = & \frac{1-y+xy}{1-4m_c^2/S},
\\
z_{2}^{min}(z_1,z_3)&=&\left(1+\frac{4m_c^2}{Q^2}\frac{z_3}{z_1}\right)
z_1xy\frac{1}{z_1z_3+y-1}
\nonumber
\end{eqnarray}

\noindent 2. Light flavour excitation:\\
In the case of the scattering off the light (massless) flavours ($u$, $d$, $s$) 
the partonic cross section is proportional to $\delta(1-\hat{x})$ and one easily
finds (of course $\Delta g \to \Delta q$, $q=u, d, s$)
\begin{equation}
\Delta \sigma_0(z_1,z_3)=
\frac{4\pi\alpha^2}{S x^2 y^2}\frac{y}{\hat{y} z_1^2}
\lambda_N \left(\hat{y}(2-\hat{y})\bar{z}_2G_1(\bar{z}_2,\hat{Q}^2)
+\hat{y}^2\bar{z}_2G_3(\bar{z}_2,\hat{Q}^2)
+(1-\hat{y})G_4(\bar{z}_2,\hat{Q}^2)
\right)
\label{mpds0}
\end{equation}
with
$\displaystyle \bar{z}_2=\frac{z_1 x y}{z_1 z_3 + y -1}$ and
the structure functions $G_i$ from eq.~(\ref{struct}) and (\ref{udsg}).
The integration bounds can be obtained from the above 
$z_{1,3}^{min}$ by 
setting $m_c \to 0$.

\subsection{$z_D$-differential case}
In this subsection we consider $z_D$-differential polarized 
heavy flavour production.
The cross sections for ISR and FSR have the same structure
as in eq.~(\ref{initial}) and (\ref{final}) with a $z_D$-dependent 
function $\Delta\sigma_0$ \cite{ijs97}:
\begin{eqnarray}
\frac{d^3\Delta\sigma^i}{dxdydz_D} & = & \frac{\alpha L_{e}}{2\pi}
\int^1_{z_{1}^{min}}dz_1\left[\frac{1+z_1^2}{1-z_1}
(\Delta\sigma_0(z_1,1;z_D)-\Delta\sigma_0(1,1;z_D))\right] \nonumber
\\
&& +\ \frac{\alpha L_{e}}{2\pi}H(z_{1}^{min})\Delta\sigma_0(1,1;z_D),
\label{fragini}\\
\frac{d^3\Delta\sigma^f}{dxdydz_D} & = & \frac{\alpha L_{e}}{2\pi}
\int^1_{z_{3}^{min}}dz_3\left[\frac{1+z_3^2}{1-z_3}
(\Delta\sigma_0(1,z_3;z_D)-\Delta\sigma_0(1,1;z_D))\right] \nonumber
\\
&& +\frac{\alpha L_{e}}{2\pi}H(z_{3}^{min})\Delta\sigma_0(1,1;z_D),
\label{fragfin}
\end{eqnarray}
with
\begin{eqnarray}
\Delta\sigma_0(z_1,z_3;z_D) & = &\int^1_{z_{2}^{min}}dz_2 \Delta g
\left(z_2,\mu ^2\right)\int_{z_{4}^{min}}^{z_{4}^{max}}dz_4
\frac{\partial{(\hat{x},\hat{y},\hat{z_c})}}{\partial{(x,y,z_D)}}
\frac{d^3\Delta\hat{\sigma}}{d\hat{x}d\hat{y}d\hat{z_c}}D_c(z_4)\ .
\label{frags0}
\end{eqnarray}
The partonic cross section is given by
\begin{eqnarray}
\frac{d^3\Delta\hat{\sigma}}{d\hat{x}d\hat{y}d\hat{z_c}}&=&
\frac{4\pi \alpha^2}{\hat{S}\hat{x}^2\hat{y}^2}\lambda_N
\Big[\hat{y}(2-\hat{y})\hat{x}
\hat{G}_1(\hat{x},\hat{Q}^2,\hat{z}_c)
+\hat{y}^2 \hat{x}\hat{G}_3(\hat{x},\hat{Q}^2,\hat{z}_c)
\nonumber\\
&&\qquad \qquad +(1-\hat{y})\hat{G}_4(\hat{x},\hat{Q}^2,\hat{z}_c)
\Big]\ ,
\label{zpartonic}
\end{eqnarray}
inserting the $z_h$-differential structure functions from eq.~(\ref{generalg})
in
\begin{eqnarray}
\hat{G}_1(\hat{x},\hat{Q}^2,\hat{z}_c)&=&\lambda\sum_{B,B^\prime} C_l^{BB^\prime}
\chi_B(\hat{Q}^2)\chi_{B^\prime}(\hat{Q}^2)
\hat{g}_1^{BB^\prime}(\hat{x},{Q}^2,\hat{z}_c)
\nonumber\\
\hat{G}_{3,4}(\hat{x},\hat{Q}^2,\hat{z}_c)&=&\sum_{B,B^\prime} C_l^{BB^\prime}
\chi_B(\hat{Q}^2)\chi_{B^\prime}(\hat{Q}^2)
g_{3,4}^{BB^\prime}(\hat{x},{Q}^2,\hat{z}_c)\ .
\label{zsubstruct}
\end{eqnarray}
The Jacobian in eq.~(\ref{frags0}) can easily be calculated as
\begin{eqnarray}
\frac{\partial{(\hat{x},\hat{y},\hat{z_c})}}{\partial{(x,y,z_D)}}=
\frac{\partial{(\hat{x},\hat{y})}}{\partial{(x,y)}}
\frac{\partial \hat{z}_c}{\partial z_D}=
\frac{y^2}{z_2z_4(z_1z_3-1+y)^2}.
\end{eqnarray}
The additional integration bounds read:
\begin{eqnarray}
z_{4}^{max}&=&\min \left[1,\frac{z_D r}{\hat{z}_{c}^{min}}\right],
\nonumber\\
z_{4}^{min}&=&\min \left[z_{4}^{max},\frac{z_D r}{\hat{z}_{c}^{max}}
\right]
\end{eqnarray}
with $r$ from eq.~(\ref{r}) and 
$\hat{z}_{c}^{min,max}=z_{h=c}^{min,max}(\hat{x},\hat{Q}^2)$ 
from eq.~(\ref{zhbound}).

For $D_c(z)$, we use a Peterson et al.\ fragmentation function 
\cite{peterson83}
\begin{equation}
D_c(z) = N \left\{ z \left[ 1-z^{-1}-\varepsilon_c/(1-z)
\right]^2\right\}^{-1}\ ,
\label{peterson}
\end{equation}
normalized to $\int_0^1 dz D_c(z) = 1$, i.~e.~,
\begin{equation}
N^{-1} =
\frac{({\varepsilon_c}^2-6{\varepsilon_c}+4)}{(4-{\varepsilon_c}) 
\sqrt{4{\varepsilon_c}-{\varepsilon_c}^2}}
\left\{
\arctan\frac{{\varepsilon_c}}{\sqrt{4{\varepsilon_c}
-{\varepsilon_c}^2}}
+ \arctan\frac{2-{\varepsilon_c}}{\sqrt{4{\varepsilon_c}
-{\varepsilon_c}^2}} \right\}
+ \frac{1}{2} \ln {\varepsilon_c} + \frac{1}{4-{\varepsilon_c}}\ .
\end{equation} 
\section{Numerical results}
The production of heavy flavours will be measured in several future
experiments, e.g., 
the charm upgrade of the HERMES experiment \cite{hermesc}, the COMPASS 
experiment \cite{compass} and a possible
polarized HERA collider mode \cite{polhera}.
In the following, we discuss the radiative corrections to polarized
heavy flavour leptoproduction in the kinematical regions of these experiments.

As usual we introduce the correction factor $\delta$ by 
$d\Delta \sigma=d\Delta\sigma^0(1+\delta)$, i.~e., 
$\delta=\delta^i+\delta^f+\ldots$
with 
\begin{displaymath}
\delta^{i,f}(x,y)=\frac{d^2\Delta \sigma^{i,f}}{dx dy}/
\frac{d^2\Delta \sigma^0}{dx dy}
\end{displaymath}
or, in the $z_D$-differential case, 
\begin{displaymath}
\delta^{i,f}(x,y,z_D)=\frac{d^3\Delta \sigma^{i,f}}{dx dy dz_D}/
\frac{d^3\Delta \sigma^0}{dx dy dz_D}.
\end{displaymath}
The labels ``i'' and ``f'' stand for ``initial'' and ``final'' state radiation,
$d\Delta \sigma^0$ denotes the Born cross section under consideration.

In the following figures we employ the polarized parton densities 
GRSV(LO) 'Standard' \cite{grsv96} with a positive definite polarized
gluon and GS(LO) 'C' \cite{gs96} with a non-positive definite gluon.
It should be stressed again that both of these different parametrizations
describe present data.
Furthermore, theoretical uncertainties due to different choices of the
charm mass $m_c$ and the factorization scale $\mu$ can be sizeable not
only in the cross sections but also in the ratio $\delta$.
In all figures the charm mass has been set to $m_c=1.5$ GeV.  
An appropriate choice of $\mu$ is not known.
In order to find a factorization scale $\mu$ leading to a good
perturbative stability affords the knowledge of the next-to-leading 
order corrections
to polarized heavy flavour production which have not yet been calculated.
Thus, we use the two factorization scales $\mu_1^2=4 m_c^2$ and
$\mu_2^2=Q^2+4 m_c^2$, which have been favoured in the 
unpolarized case \cite{scale}.
For completeness, all figures have been calculated with $\lambda=\lambda_N=1$.
\subsection{HERMESC}
\begin{figure}[th]
\centering

\epsfig{figure=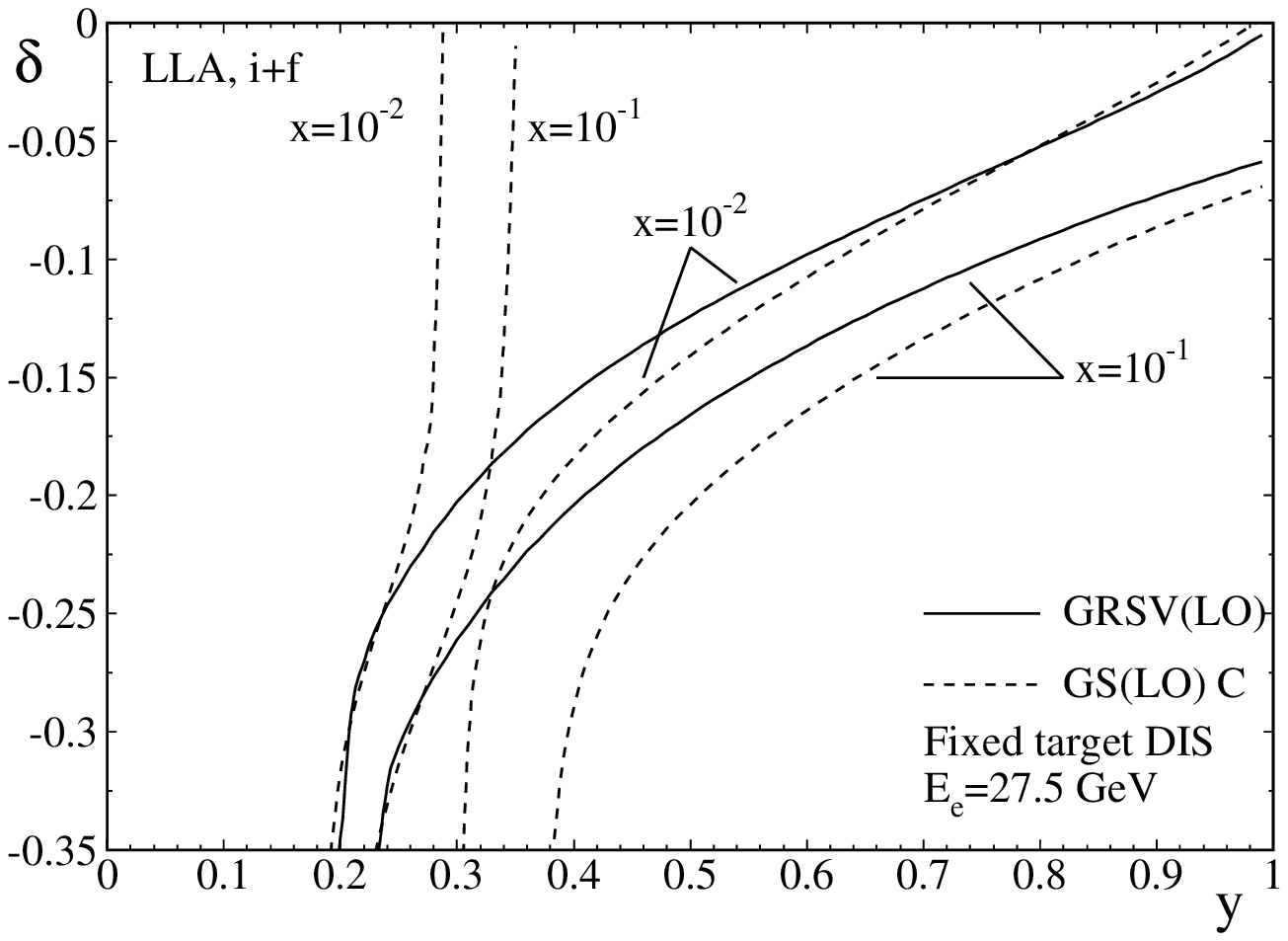,width=14.0cm}
\caption{\sf Radiative corrections to longitudinally polarized heavy quark 
electroproduction in $\cal{O}(\alpha)$-LLA at $\sqrt{S}=7.4$ GeV (HERMES).}
\label{hermesc}
\end{figure}

The detector upgrade of the HERMES
experiment \cite{hermesc} is dedicated to measure polarized leptoproduction
of open charm via the scattering of
longitudinally polarized electrons ($E_e=27.5$ GeV) off a longitudinally 
polarized 
fixed nucleon target, i.~e.~, at $\sqrt{S}=7.4$ GeV. 
In fig.~\ref{hermesc} 
[$m_c=1.5$ GeV, $\mu^2=Q^2+4m_c^2\approx 4m_c^2$, parton distributions:
GRSV(LO) 'Standard' (solid line), GS(LO) 'C' (dashed line)] 
we show the radiative corrections in $\cal{O}(\alpha)$-LLA to this
process according to eq.~(\ref{initial})--(\ref{pgfs0}) for two values of 
Bjorken-$x$ in dependence of $y$.
From the threshold condition $\hat{s}=(p_c+p_{\bar{c}})^2 \ge 4 m_c^2$ follows
$y \ge 4 m_c^2/(S(1-x))=:y_{min}\approx 0.16/(1-x)$.
In the range $y_{min}\le y \le 0.4$ the corrections are big and strongly
depend on the parametrization of the polarized gluon.
In the case of the GS parton distributions the correction factor $\delta$ 
diverges at $y \approx 0.30$ for
$x=10^{-2}$ and $y \approx 0.35$ for $x=10^{-1}$ due to a change of sign of
 the Born cross section in the denominator. 
However, in this range the cross section $d\Delta \sigma$ is small. 
In the limit $y \to y_{min}$ $d\Delta \sigma$ vanishes.
In the physically relevant region $y\gtrsim 0.5$ the corrections are
smaller than $20 \%$ further decreasing with increasing $y$.
Uncertainties due to the parton distributions are
smaller than $5 \%$. Using a factorization scale $\mu^2=\hat{s}$ leads to
negligible changes in figure \ref{hermesc}.
\subsection{COMPASS}
\begin{figure}[th]
\centering

\epsfig{figure=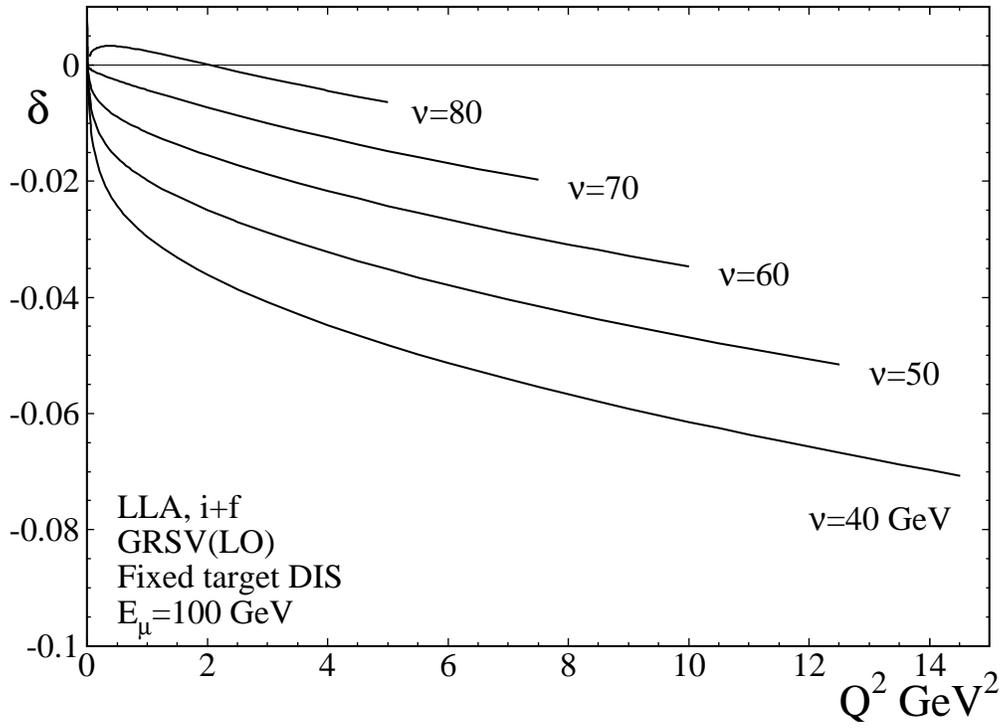,width=14.0cm}
\caption{\sf Radiative corrections to longitudinally polarized heavy quark 
muoproduction in $\cal{O}(\alpha)$-LLA at $\sqrt{S}=14.1$ GeV (COMPASS).}
\label{compassc}
\end{figure}

The COMPASS experiment \cite{compass}
is going to measure polarized open charm muoproduction at an energy 
of $\sqrt{S}=14.1$ GeV ($E_{\mu}=100$ GeV). 
The radiative corrections for muoproduction can be obtained by simply
replacing $L_e=\ln Q^2/m_e^2 \to \ln Q^2/m_{\mu}^2=L_{\mu}$.
Since nothing else is changed the corrections $\delta(x,y)$ are expected to 
be similar to fig.~\ref{hermesc} but suppressed by the factor $L_{\mu}/L_e<0.5$.
Fig.~\ref{compassc} 
[$m_c=1.5$ GeV, $\mu^2=Q^2+4m_c^2\approx 4m_c^2$, parton distribution:
GRSV(LO) 'Standard'] 
shows the correction factor $\delta(Q^2,\nu)$ ($y=\nu/E_{\mu}$, $x=Q^2/2M\nu$)
for the ($\nu$, $Q^2$)-range given in table 3.1 of \cite{compass}.
Only for $\nu=40$ GeV and $Q^2\ge 6$ GeV$^2$ the corrections exceed $5 \%$.
\subsection{Polarized HERA collider mode}
In a polarized HERA collider  mode longitudinally polarized electrons 
are scattered off longitudinally
polarized
protons with a center-of-mass energy $\sqrt{S}=300$ GeV.
One of the main goals of a polarized HERA is the measurement of the
spin dependent structure function $g_1(x,Q^2)$ at values of ($x$, $Q^2$)
which are not accessible to other experiments \cite{ball}.
In the range of small $x\lesssim 10^{-3}$ the charm contribution $g_1^c(x,Q^2)$
to $g_1^P(x,Q^2)$ has to be accounted for.

\begin{figure}[h]
\centering
\epsfig{figure=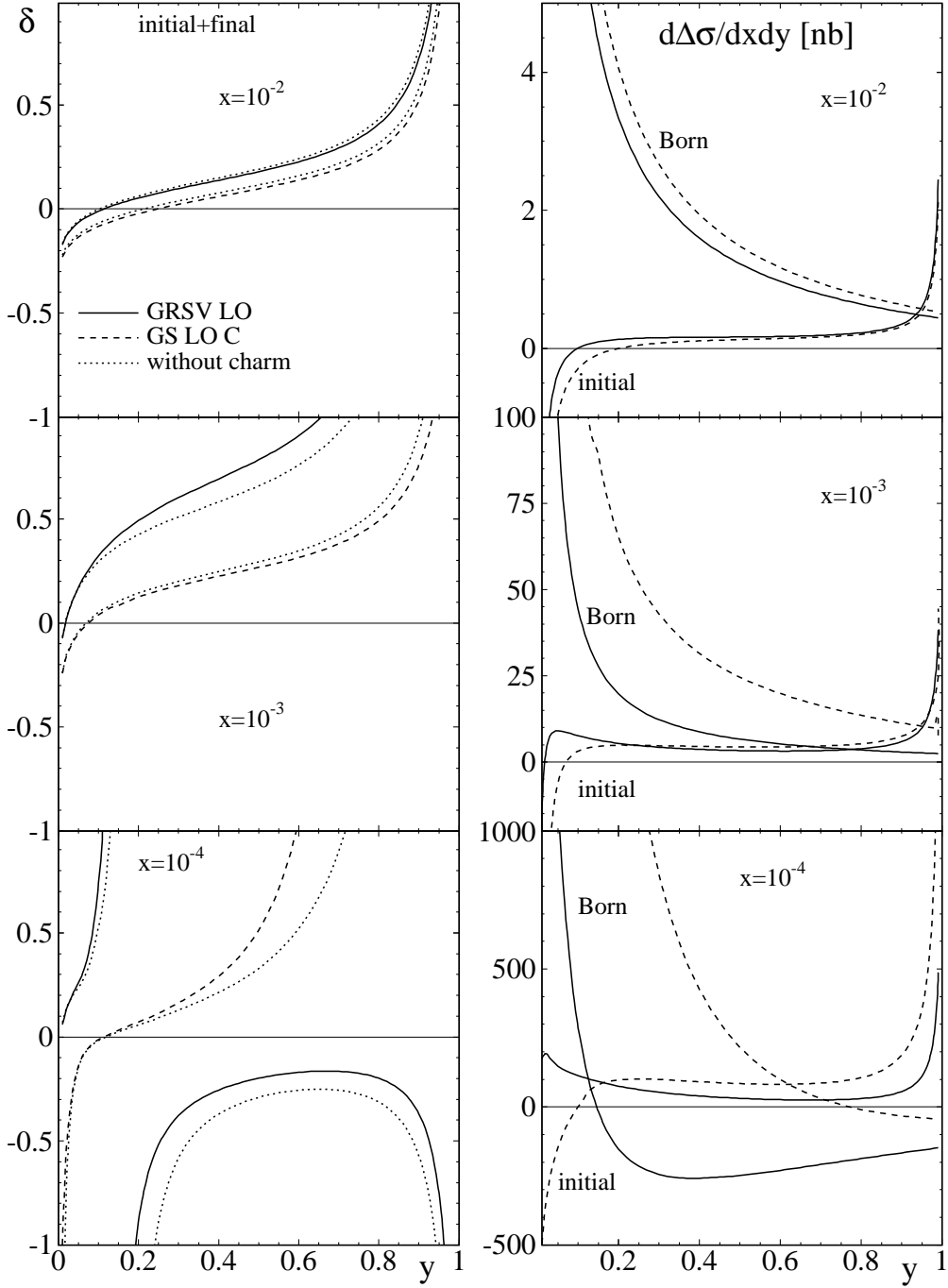,width=14cm}
\caption{\sf Right side: $d\Delta\sigma^{0,i}(ep \to eX)$ 
(u,d,s and heavy (charm) quark contribution) at $\sqrt{S}=300$ GeV for
$x=10^{-2}$, $10^{-3}$, and $10^{-4}$, using the
GRSV(LO) 'S' \protect\cite{grsv96} (solid line) and the GS(LO) 'C' 
\protect\cite{gs96}(dashed line)
parton densities and a factorization scale $\mu^2=Q^2+4m_c^2$ with $m_c=1.5$ GeV.
Left side: The corresponding correction factor $\delta(x,y)$.
The dotted lines have been calculated without heavy (charm) quark contribution.}
\label{rcudsc4}
\end{figure}

In fig.~\ref{rcudsc4} the charm contribution to the radiative corrections
to $d^2\Delta \sigma/dx dy(ep \to eX)$ is illustrated.
On the right side we show the electromagnetic (em) cross sections 
$d\Delta\sigma^{0,i}(ep \to eX)$, including the $u$, $d$, $s$ contributions 
from electron quark scattering (eq.~(\ref{frags0})) and 
the charm contribution due to the 
photon gluon fusion (eq.~(\ref{pgfs0})), 
at $\sqrt{S}=300$ GeV for
$x=10^{-2}$, $10^{-3}$, and $10^{-4}$, using the
GRSV(LO) 'S' \cite{grsv96} (solid line) and the GS(LO) 'C' \cite{gs96} 
(dashed line)
parton densities and a factorization scale 
$\mu^2=Q^2+4m_c^2$ with $m_c=1.5$ GeV (for the charm contribution).
The cross sections for FSR have similar shapes as for ISR.
The corresponding correction factors $\delta(x,y)$ can be seen on the left hand 
side.
The dotted lines have been calculated without heavy (charm) quark contribution.
Employing the GRSV parton distributions
they agree with the dashed lines in the figures 11 and 12 in \cite{bb96}.
However, for $x\lesssim 10^{-2}$ the corrections strongly depend on the 
choice of the parametrization of the polarized parton densities as can be
seen by comparison of the solid with the dashed lines.
This point has not been hinted at in the literature.
The charm contribution to the radiative corrections is roughly $10 \%$
in regions where the corrections do not exceed $50 \%$ and even bigger
otherwise.

\begin{figure}[ht]
\centering
\epsfig{figure=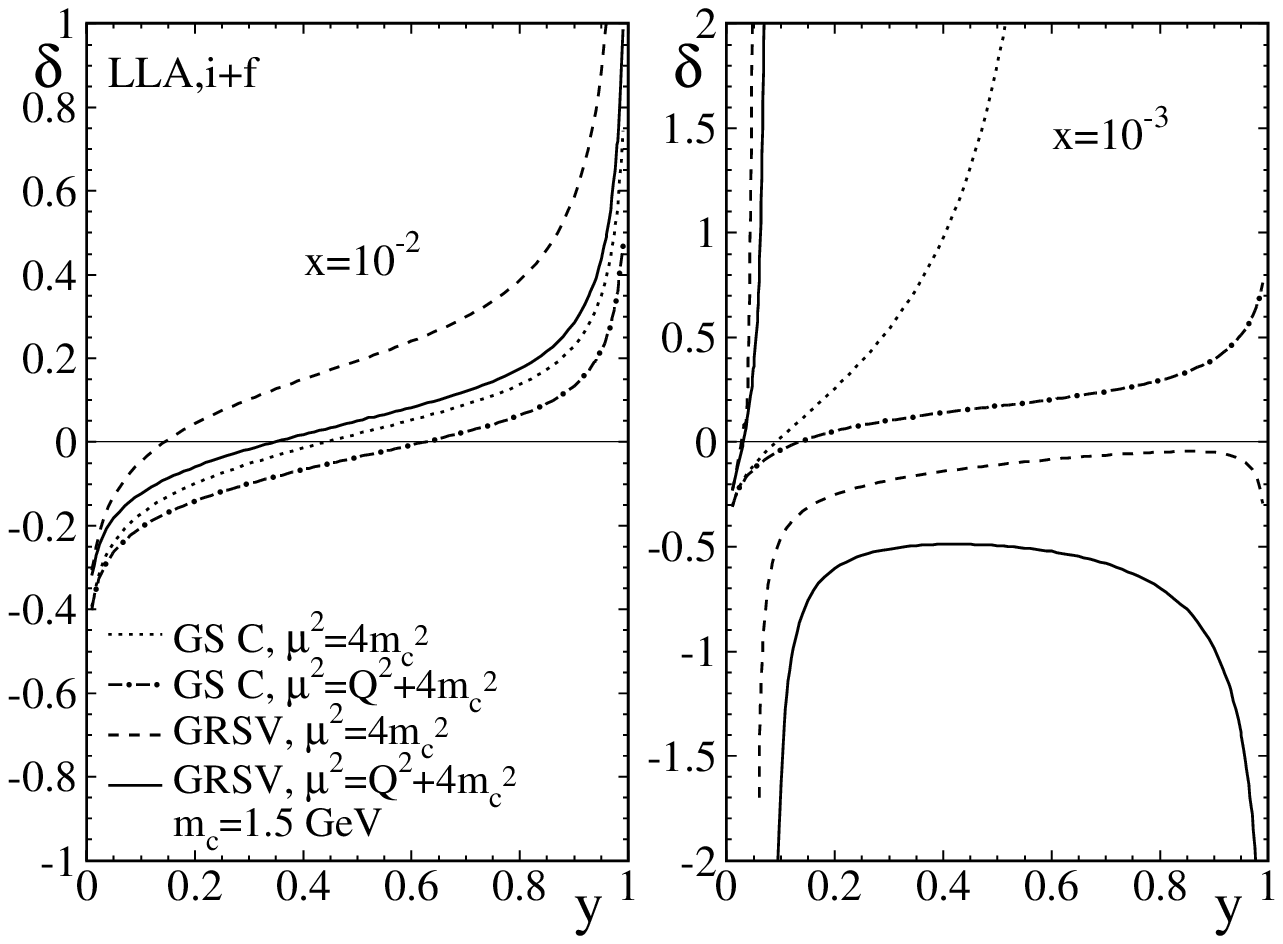,width=12cm}
\caption{\sf Radiative corrections to longitudinally polarized heavy quark 
electroproduction in $\cal{O}(\alpha)$-LLA at $\sqrt{S}=300$ GeV (HERA).}
\label{rcc}
\end{figure}

Similar to the unpolarized case ($F_2^c$) \cite{adloff96,zeus97} 
it would be interesting, albeit very difficult, 
to measure $g_1^c$ in order to constrain the polarized gluon 
density $\Delta g(x,Q^2)$ in the proton at small $x$.
Fig.~\ref{rcc} shows the (inclusive) radiative corrections to polarized charm
production via the photon gluon fusion.
$Z$-exchange has been neglected, since
$\hat{g}_3^c=\hat{g}_4^c=0$ (inclusive case, see eq.~(\ref{g3sub})) and 
$\hat{g}_1^{c,\gamma\gamma}\approx \hat{g}_1^{c}$.
In contrast to unpolarized heavy flavour production \cite{ijs97}
the theoretical uncertainties are not under control.
The use of the two factorization scales $\mu_1^2=4 m_c^2$ and
$\mu_2^2=Q^2+4 m_c^2$ \cite{scale}
and the parton distributions GRSV(LO) 'S' and GS(LO) 'C'  
in each case results in rather different corrections, especially for $x=10^{-3}$.
This behaviour is not only due to the unknown NLO corrections but mainly
due to the almost entirely unconstrained $\Delta g(x,Q^2)$.

\begin{figure}[ht]
\centering
\epsfig{figure=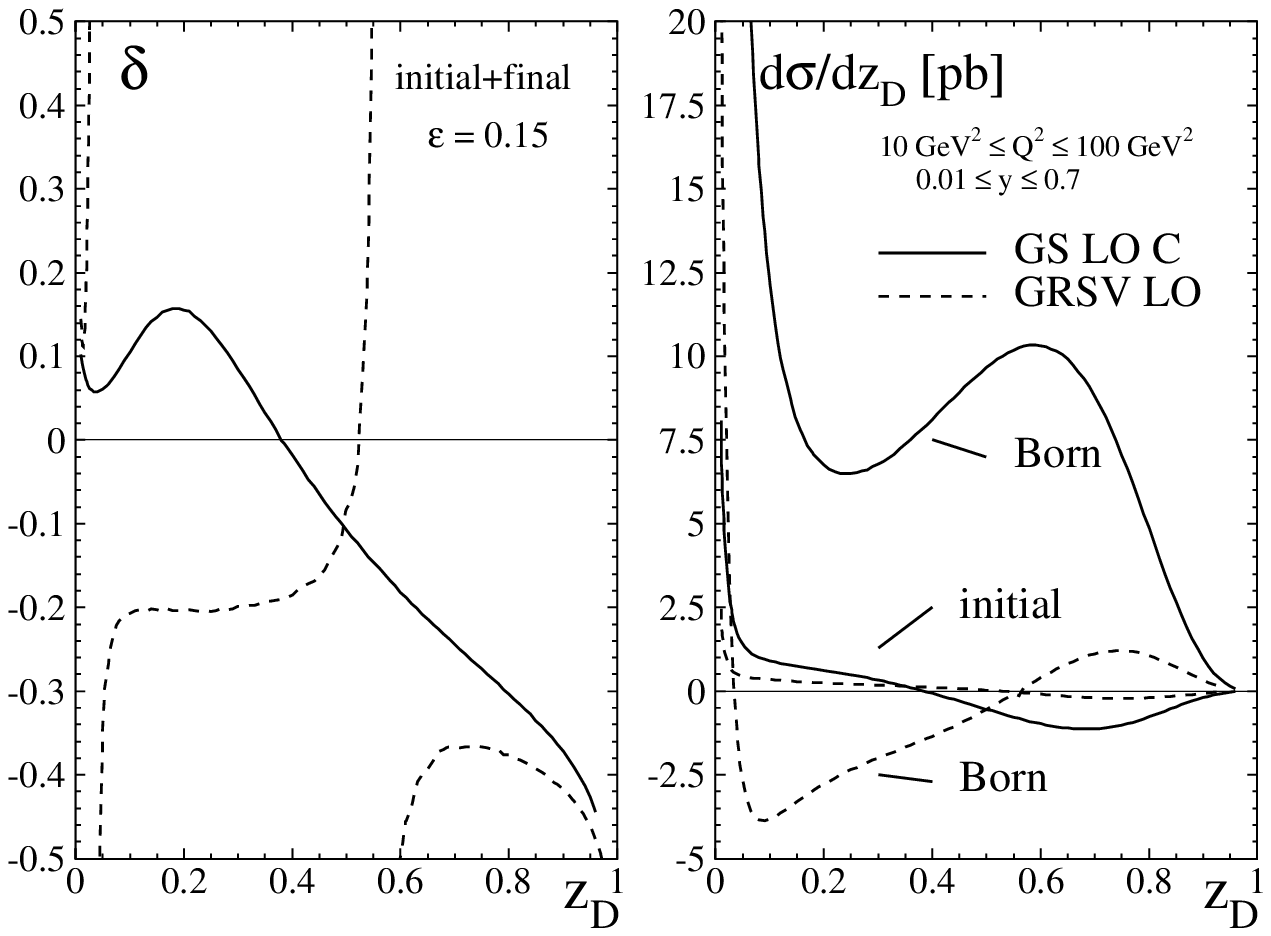,width=12cm}
\caption{\sf $z_D$-differential radiative corrections to 
deep inelastic charm production ($m_c=1.5\ {\rm GeV}$, $\mu^2=Q^2+4m_c^2$)
, integrated over the kinematical range
$10\ {\rm GeV}^2\le Q^2\le 100\ {\rm GeV}^2,\ 0.01 \le y\le 0.7$, 
using the parton distributions GRSV(LO) 'S' \protect\cite{grsv96} (solid line) 
and GS(LO) C \protect\cite{gs96} (dashed line)
and a Peterson et al.\ fragmentation function \protect\cite{peterson83}
with $\varepsilon=0.15$.}
\label{polz}
\end{figure}

As has been already stressed in \cite{ijs97} measurements of 
heavy quark production processes (one hadron inclusive processes) are of course
differential in the momentum of the (observed) heavy quark, 
recommending to perform the radiative corrections on the same
differential level.
Thus, we have repeated the semi-inclusive $z_D$-differential 
analysis of our study of unpolarized heavy quark production.
Contrary to the inclusive case, 
$g_3$ and $g_4$ do not vanish (see eq.~(\ref{generalg})) and 
should not be neglected.
Allthough $g_4(x,Q^2)\approx \chi_Z(Q^2) g_4^{\gamma Z}(x,Q^2)$ is suppressed 
at small $Q^2 \ll M_Z^2$ by the factor 
$\displaystyle \chi_Z(Q^2) \approx 0.7 \frac{Q^2}{Q^2+M_Z^2}$
its weight relativ to $g_1(x,Q^2)$ in the cross section is 
(up to terms of the order ${\cal{O}}(y^2)$) 
\begin{equation}
\frac{(1-y)g_4(x,Q^2)}{2 x y g1(x,Q^2)}
\approx \frac{\chi_Z}{2 x y} \frac{g_4^{\gamma Z}}{g1}
\end{equation}
due to the
structure of the hadronic tensor.
The same holds for the radiative corrections where the corresponding factor
in $\Delta \sigma_0(z_1,z_3)$ reads
$\displaystyle \frac{(1-\hat{y})g_4(\bar{z}_2,\hat{Q}^2)}
{2 x y g_1(\bar{z}_2,\hat{Q}^2)}$.
The factor $\chi_Z/2 x y\approx 0.7 S x y/2 x y\ \cdot 1/(Q^2+M_Z^2)$ is 
varying between approximately $0.7$ ($Q^2=S=10^5$ GeV$^2$) and 
$7$ ($Q^2 \ll M_Z^2$ (small $x$)). 
For this reason it is (a priori) not allowed to disregard $g_4$ 
(,i.e.~, contributions from $Z$-exchange)\footnote{The importance of the 
electroweak contributions
can be seen in fig.~12 of \protect\cite{bb96}.We can confirm this result and
are grateful to D. Bardin for discussions on this point.}. 
However, in the case of $z_D$-differential heavy flavour production the
differences between em and ew radiative corrections turn out to be negligible.
This could be expected because the electroweak effects from $g_{3,4}$ have to
vanish after $z_D$-integration (see eq.~(\ref{g3sub}).
In fig.~\ref{polz} we show the $z_D$-differential radiative 
corrections according to eq.~(\ref{fragini})--(\ref{frags0}) after 
additional integration
over the kinematical range 
$10\ {\rm GeV}^2\le Q^2\le 100\ {\rm GeV}^2,\ 0.01 \le y\le 0.7$.
The left hand side shows the correction factor 
$\delta(z_D)$ (initial and final state radiation), employing the
GRSV(LO) 'S' \cite{grsv96} (solid line) and the GS(LO) 'C' \cite{gs96} 
(dashed line) parton distributions.
Again we have chosen $\mu^2=Q^2+4m_c^2$ with $m_c=1.5$ GeV.
For $D_c$ a Peterson et al.\ fragmentation function \cite{peterson83} 
with $\varepsilon=0.15$ has been taken.
On the right side of fig.~\ref{polz} the $z_D$-dependence of the 
cross sections $d\Delta \sigma^{0,i}$ is displayed.
As in the inclusive case the ISR and FSR cross sections
have similar shapes.
Apart from irrelevant regions of the phase space where the Born cross section
vanishes the corrections amount between $20 \%$ and $- 40 \%$.
Applying suitable (depending on the parton distribution) cuts, e.~g.~, 
$0.2 \le z_D\le 0.8$ it is possible to further reduce the corrections as can be seen
from the rhs of fig.~\ref{polz}. 

To conclude, at small $x$ it is (up to now) only possible to perform calculations 
of radiative corrections to charm production modulo the
choice of the parametrization of the polarized gluon.
The situation will improve as soon as the gluon distribution becomes more
constrained (for $x\lesssim 10^{-2}$) and the NLO expressions for 
polarized heavy flavour production are calculated.
\section{Summary}
We have calculated fully massive QED corrections to polarized 
heavy flavour leptoproduction in ${\cal O}(\alpha)$-LLA
for HERMES, COMPASS, and HERA kinematics.
The radiative corrections to the first two experiments are smaller than
$20 \%$ and $5 \%$ respectively in relevant realms of the phase space and the 
theoretical uncertainties due to different choices of the parton distributions 
and the factorization scale are under control.
For HERA kinematics the situation changes if $x < 10^{-2}$. In this case
all results strongly depend on the chosen set of parton distributions. 
However, for any fixed parton distribution we find the charm contribution to the 
corrections to a measurement of $g_1^P(x,Q^2)$ non-negligible 
(for $x\lesssim 10^{-3}$).
Finally, we have discussed radiative corrections to $g_1^c(x,Q^2)$ 
at a polarized HERA in analogy to $F_2^c(x,Q^2)$ 
\cite{adloff96,zeus97} in the unpolarized case.
Besides the inclusive case, 
we have derived general $z_h$-differential expressions for the polarized boson
gluon fusion process. These formulae have been used to study
$z_D$-differential corrections,
keeping in mind that heavy flavour production is a semi-inclusive process
recommending to calculate the corrections on the same differential level.
These corrections turned out to lie between $20 \%$ and $- 40 \%$ 
apart from regions of the phase space where the Born cross section is small.

\section*{Acknowledgements}
It is a pleasure to thank E.\ Reya and M.\ Gl\"{u}ck for advice and 
useful discussions.
This work has been supported by the
Graduiertenkolleg 'Erzeugung und Zerf\"alle von Elementarteilchen' of the 
University of Dortmund.

\newpage
%
%
%
%
%

\newpage
\begin{thebibliography}{99}
%
\bibitem{gehrmann97}
T.~Gehrmann, Summary of the Working Group IV ``Spin Physics'' of the
International Workshop on ``Deep Inelastic Scattering and QCD'' 
(DIS '97), hep-ph/9706351;
\\
M.~Stratmann, in Proceedings of the 2nd Topical Workshop on 'DIS off
Polarized Targets: Theory Meets Experiment', Zeuthen, Germany, 1-5 Sep 1997
(hep-ph/9710379).
\bibitem{hermesc}
The HERMES Collab., The HERMES Charm Upgrade Program, HERMES 97-004.
\bibitem{compass}
G.~Baum et al., COMPASS Collab., CERN/SPSLC 96-14.
\bibitem{polhera}
J.~Feltesse and A.~Sch\"afer, in \emph{Proceedings of the workshop on 
Future Physics at HERA}, DESY, 1996, G.~Ingelman, 
A.~De Roeck and R.~Klanner (ed.) vol. 2, p. 760.
\bibitem{ball}
R.~Ball et al., in \emph{Proceedings of the workshop on 
Future Physics at HERA}, DESY, 1996, G.~Ingelman, 
A.~De Roeck and R.~Klanner (ed.) vol. 2, p. 777.
\bibitem{bb96}
D.~Bardin, J.~Bl\"umlein, P.~Christova, and L.~Kalinovskaya, 
Nucl. Phys. \textbf{B506} (1997) 295.
\bibitem{ijs97}
I.~Schienbein, DO-TH 97/05, hep-ph/9705322, Eur. Phys. J. \textbf{C}(to appear), 
DOI 10.1007/s100529800714.
\bibitem{ais97}
I.~Akushevich, A.~Ilyichev, and N.~Shumeiko, hep-ph/9710486.
\bibitem{bk}
J.~Bl\"umlein and N.~Kochelev, Phys. Lett. \textbf{B381} (1996) 296;
Nucl. Phys. \textbf{B498} (1997) 285.
\bibitem{manohar}
A.~V.~Manohar, An Introduction to spin dependent deep inelastic scattering.
Lectures given at Lake Louise Winter Inst., Lake Louise, Canada, Feb 23-29, 1992,
hep-ph/9204208.
\bibitem{vw91}
W.~Vogelsang and A.~Weber, Nucl. Phys. \textbf{B362} (1991) 3.
\bibitem{ks97}
S.~Kretzer and I.~Schienbein, Phys. Rev. \textbf{D56} (1997) 1804.
\bibitem{lla}
See, for example:
J.~Bl{\"{u}}mlein, Z. Phys. \textbf{C47} (1990) 89;
\\
J.~Kripfganz, H.-J.~M{\"o}hring, and H.~Spiesberger, 
Z. Phys. \textbf{C49} (1991) 501.
\bibitem{schuler88}
G.~Schuler, Nucl. Phys. \textbf{B299} (1988) 21.
\bibitem{peterson83}
C.~Peterson et al., Phys. Rev. \textbf{D27} (1983) 105.
\bibitem{grsv96}
M.~Gl\"uck, E.~Reya, M.~Stratmann, and W.~Vogelsang, 
Phys. Rev. \textbf{D53} (1996) 4775.
\bibitem{gs96}
T.~Gehrmann and W.~J.~Stirling, Phys. Rev. \textbf{D53} (1996) 6100.
\bibitem{scale}
M.~Gl{\"{u}}ck, E.~Reya, and M.~Stratmann, 
Nucl. Phys. \textbf{B422} (1994) 37;
\\ 
A.~Vogt, DESY 96-012, hep-ph/9601352.
\bibitem{adloff96}
{C.~Adloff et al., H1 Collab.}, Z. Phys. \textbf{C72} (1996) 593. 
\bibitem{zeus97}
{J.~Breitweg et al., ZEUS Collab.}, Phys. Lett. \textbf{B407} (1997) 402.

\end{thebibliography}
\end{document}